\documentclass[preprint,nofootinbib,amsmath,amssymb,preprintnumbers]{revtex4}

\usepackage{bm}% bold math
\usepackage[colorlinks=true,linkcolor=blue,citecolor=blue]{hyperref}
\usepackage{graphicx}

\newcommand{\be}{\begin{equation}}
\newcommand{\ee}{\end{equation}}

\begin{document}

\preprint{MIT-CTP/4160}

\title{Effective Action, Boundary Conditions, and Virasoro Algebra for AdS$_3$}

\author{Achilleas P. Porfyriadis and Frank Wilczek\\
\small\it Center for Theoretical Physics, MIT\\   \small\it  Cambridge MA 02139 USA}

\date{\today}

\begin{abstract}
We construct, to second order, the effective action of General Relativity for small excitations generated by a vector field and use it to study conformal symmetry in the boundary of AdS$_3$. By requiring finiteness of the boundary effective action(s) for certain asymptotic transformations, we derive the well-known Virasoro algebra and central charge associated with the boundary of AdS$_3$.  The bulk action for these transformations can be arbitrarily small.
\end{abstract}

\maketitle

\section{Introduction and Motivation}

Brown and Henneaux \cite{Brown:1986nw} studied asymptotic symmetries of three-dimensional Anti-de-Sitter space (AdS$_3$),
\be\label{AdS3}
ds^2=-\left(1+\frac{r^2}{l^2}\right)dt^2+\left(1+\frac{r^2}{l^2}\right)^{-1}dr^2+r^2d\phi^2\,.
\ee
They found that these symmetries contain two copies of the Virasoro algebra with central charge $c=3l/(2G)$, where $l$ is the AdS$_3$ radius and $G$ is the three dimensional Newton constant.
Strominger \cite{Strominger:1997eq} used this result to shed light on the microscopic origin of the Bekenstein-Hawking entropy of BTZ \cite{Banados:1992wn, Banados:1992gq} (and related) black holes.   Specifically, he  invoked Cardy's formula that connects central charge to the density of states in conformal field theory, with $c=3l/(2G)$, and found numerical agreement with the usual, semiclassical, Bekenstein-Hawking entropy.    This suggests that \emph{any} consistent quantum theory of gravity on AdS$_3$ is dual to a two-dimensional conformal field theory (CFT) with $c=3l/(2G)$.
Recently these results were extended to extremal Kerr (and related) black holes, in what is known as the Kerr/CFT correspondence \cite{Guica:2008mu}. In Kerr/CFT, asymptotic symmetries of the Near-Horizon-Extremal-Kerr (NHEK) metric \cite{Bardeen:1999px} form one copy of the Virasoro algebra with central charge $c=12J$ (where $J$ the angular momentum).   Here too, the Cardy formula reproduces the Bekenstein-Hawking entropy of the  black hole.

In these analyses, the definition of asymptotic symmetry depends sensitively on precise specification of fall-off conditions at infinity.  Henneaux and Teitelboim \cite{Henneaux:1985tv} proposed some general requirements for the boundary conditions: (i) they should be invariant under the AdS$_3$ isometry group, (ii) they should allow for the asymptotically AdS$_3$ solutions of physical interest (the BTZ black hole here), and (iii) they should yield finite charges in the canonical (i.e. Hamiltonian) formalism of General Relativity (GR). For example, Brown-Henneaux chose:
\begin{equation}\label{asymptotically AdS3 with single O}
\begin{gathered}
g_{tt}=-\frac{r^2}{l^2}+{\cal O}(1)\,, \quad g_{tr}={\cal O}(\frac{1}{r^3})\,,\quad g_{t\phi}={\cal O}(1)\,,\\
g_{rr}=\frac{l^2}{r^2}+{\cal O}(\frac{1}{r^4})\,,\quad g_{r\phi}={\cal O}(\frac{1}{r^3})\,,\quad g_{\phi\phi}=r^2+{\cal O}(1)\,.
\end{gathered}
\end{equation}
The corresponding asymptotic symmetries are then given by:
\begin{eqnarray}\label{asymptotic isometry vectors with single O}
\xi^t&=&l\left(T^++T^-\right)+\frac{l^3}{2r^2}\left({T^+}''+{T^-}''\right)+{\cal O}(\frac{1}{r^4})\,,\nonumber\\
\xi^r&=&-r\left({T^+}'+{T^-}'\right)+{\cal O}(\frac{1}{r})\,,\\
\xi^\phi&=&T^+-T^--\frac{l^2}{2r^2}\left({T^+}''-{T^-}''\right)+{\cal O}(\frac{1}{r^4})\,,\nonumber
\end{eqnarray}
where $T^+=T^+(x^+)$ and $T^-=T^-(x^-)$ are arbitrary functions of a single argument $x^{\pm}=t/l\pm\phi$. One can check that (\ref{asymptotically AdS3 with single O}--\ref{asymptotic isometry vectors with single O}) meet the requirements (i-iii). However, the consistency conditions (i-iii) do \emph{not} fix the boundary conditions uniquely.  For example, one can chose more relaxed boundary conditions:
\begin{equation}\label{asymptotically AdS3 with relaxed Otr and Orphi}
\begin{gathered}
g_{tt}=-\frac{r^2}{l^2}+{\cal O}(1)\,, \quad g_{tr}={\cal O}(\frac{1}{r^2})\,,\quad g_{t\phi}={\cal O}(1)\,,\\
g_{rr}=\frac{l^2}{r^2}+{\cal O}(\frac{1}{r^4})\,,\quad g_{r\phi}={\cal O}(\frac{1}{r^2})\,,\quad g_{\phi\phi}=r^2+{\cal O}(1)\,,
\end{gathered}
\end{equation}
for which the corresponding asymptotic symmetries are given by:
\begin{eqnarray}\label{asymptotic isometry vectors with relaxed Otr and Orphi}
\xi^t&=&l\left(T^++T^-\right)+\frac{l^3}{2r^2}\left({T^+}''+{T^-}''\right)+{\cal O}(\frac{1}{r^3})\,,\nonumber\\
\xi^r&=&-r\left({T^+}'+{T^-}'\right)+{\cal O}(\frac{1}{r})\,,\\
\xi^\phi&=&T^+-T^--\frac{l^2}{2r^2}\left({T^+}''-{T^-}''\right)+{\cal O}(\frac{1}{r^3})\,.\nonumber
\end{eqnarray}
The requirements (i-ii) are met by the relaxed set (\ref{asymptotically AdS3 with relaxed Otr and Orphi}--\ref{asymptotic isometry vectors with relaxed Otr and Orphi}), since both the metric \eqref{asymptotically AdS3 with relaxed Otr and Orphi} and the asymptotic Killing vectors \eqref{asymptotic isometry vectors with relaxed Otr and Orphi} are relaxed versions of Brown-Henneaux's \eqref{asymptotically AdS3 with single O} and \eqref{asymptotic isometry vectors with single O} respectively. As for condition (iii), one can check that the charges $Q[\xi]$ corresponding to the relaxed set (\ref{asymptotically AdS3 with relaxed Otr and Orphi}--\ref{asymptotic isometry vectors with relaxed Otr and Orphi}) are still finite and equal to the ones for (\ref{asymptotically AdS3 with single O}--\ref{asymptotic isometry vectors with single O}). In both cases, expanding into modes $T^{\pm}\sim\frac{1}{2}e^{inx^\pm}$ one finds that to leading order in $1/r$, the space-time generators $\xi_n^{\pm}$ form under Lie brackets two copies of the classical centerless Virasoro algebra:
\be\label{centerless Virasoro}
[\xi_m^{\pm},\xi_n^{\pm}]_{L.B.}=i(n-m)\xi_{m+n}^{\pm}\,.
\ee
To leading order in $1/r$, the Poisson bracket algebra of the canonical generators is isomorphic to the Lie bracket algebra of the corresponding space-time generators up to a possible central extension which can be evaluated at the Dirac bracket level at $t=0$ on AdS$_3$.  In units where $16\pi G=1$, that evaluation yields:
\be\label{Dirac bracket algebra}
\{Q[\xi_m^{\pm}],Q[\xi_n^{\pm}]\}_{D.B.}=i(n-m)Q[\xi_{m+n}^{\pm}]+2i\pi ln(n^2-1)\delta_{m+n,0}\,.
\ee
With $L_m^{\pm}\equiv Q[\xi_m^{\pm}]$, upon passing from the Dirac bracket to the commutator $\{\,,\}_{D.B.}\to-i[\,,]$ we get the quantum Virasoro algebra with central charge $c=3l/(2G)$ (restoring $G$).

To every consistent set of boundary conditions, such as \eqref{asymptotically AdS3 with single O} or \eqref{asymptotically AdS3 with relaxed Otr and Orphi}, corresponds an asymptotic symmetry group (ASG) which is defined as the set of allowed symmetry transformations, such as \eqref{asymptotic isometry vectors with single O} and \eqref{asymptotic isometry vectors with relaxed Otr and Orphi} respectively, modulo the set of trivial symmetry transformations.  Here ``trivial'' means that there is no associated charge and that the canonical generator vanishes upon imposing the constraints. Therefore, given \eqref{Dirac bracket algebra}, the ASG corresponding to both sets of boundary conditions, \eqref{asymptotically AdS3 with single O} and \eqref{asymptotically AdS3 with relaxed Otr and Orphi}, is the conformal group in two dimensions with $c=3l/(2G)$.

Since the emergence of the Virasoro algebra seems depends crucially on the choice of boundary conditions, one would like to have a convincing argument justifying that choice. In this Letter, we will derive Virasoro algebra generating vectors independently by requiring finiteness of the boundary effective action(s) for certain asymptotic transformations.  We will see that these transformations indeed happen to be asymptotic symmetries of new asymptotically AdS$_3$ spaces with further relaxed but still consistent boundary conditions.  However, we also wish to stress that we do not impose boundary conditions and an ASG {\it a priori}. The emergence of the Virasoro algebra in the asymptotics of AdS$_3$ is independent of any assumed ASG and our derivation of the algebra is grounded in the physical requirement of finite action.

\section{Effective action for General Relativity}

Our goal is to construct (to second order) the effective action of GR for small excitations $g_{\mu\nu}\to g_{\mu\nu}+{\cal L}_\xi g_{\mu\nu}$ and derive the corresponding equation of motion. We start from the Einstein-Hilbert action of GR,
\be\label{Einstein-Hilbert action}
S=\int_M d^n x\,\sqrt{-g}\,(R-2\Lambda)\,.
\ee
Putting $g_{\mu\nu}\to g_{\mu\nu}+h_{\mu\nu}$, we expand to second order in $h$:
\be\label{S0[h]+S1[h]+S2[h]}
S=S^{(0)}[h]+S^{(1)}[h]+S^{(2)}[h]+{\cal O}(h^3)\,.
\ee
We then put $h_{\mu\nu}={\cal L}_\xi g_{\mu\nu}=\nabla_{\mu}\xi_{\nu}+\nabla_{\nu}\xi_{\mu}$ and obtain the desired effective action for $\xi$. The first order action for $\xi$ reads:
\begin{eqnarray}\label{1st order action for ksi}
S^{(1)}[\xi]&=&-2\int_M d^n x\,\sqrt{-g}\,(G^{\mu\nu}+\Lambda g^{\mu\nu})\nabla_{\mu}\xi_{\nu}\nonumber\\
&&+\int_{\partial M} d^{n-1} x\,\sqrt{-\gamma}\,n^{\mu} (\Box\,\xi_{\mu}+\nabla^{\nu}\nabla_{\mu}\xi_{\nu} -2\nabla_{\mu}\nabla_{\sigma}\xi^{\sigma})\,.
\end{eqnarray}
Upon integrating  the bulk piece above by parts, and using the contracted Bianchi identity, $\nabla^{\mu}G_{\mu\nu}=0$, we find that the first order action \eqref{1st order action for ksi} reduces to a boundary term:
\begin{eqnarray}\label{boundary 1st order action for ksi}
S^{(1)}[\xi]&=&\int_{\partial M} d^{n-1} x\,\sqrt{-\gamma}\,n^{\mu} \left(\Box\,\xi_{\mu}-\nabla_{\nu}\nabla_{\mu}\xi^{\nu}+(R-2\Lambda)\xi_{\mu}\right)\,.
\end{eqnarray}
Thus at first order the effective action for $\xi$ is just a boundary term. Note that Bianchi identity is an identity on curvature and so the first order action \eqref{boundary 1st order action for ksi} is obtained \emph{without} assuming the Einstein Field Equations (EFE) for the background $g_{\mu\nu}$. This is a consequence of the diffeomorphism invariance of the Einstein-Hilbert action: the transformation $g_{\mu\nu}\to g_{\mu\nu}+\nabla_{\mu}\xi_{\nu}+\nabla_{\nu}\xi_{\mu}$ is an infinitesimal diffeomorphism (\emph{i.e.} a diffeomorphism to first order in $\xi$) and therefore, at first order, the Einstein-Hilbert action changes only by a boundary term.

The second order action for $\xi$, after much simplification on the bulk term, takes the form:
\begin{eqnarray}\label{2nd order action for ksi --good writing}
S^{(2)}[\xi]&=&-\int_M d^n x\,\sqrt{-g}\,(G^{\mu\nu}+\Lambda g^{\mu\nu})(\xi_{\alpha;\mu\nu}-R_{\alpha\mu\nu\sigma}\xi^{\sigma})\,\xi^\alpha\nonumber\\
&&-\int_{\partial M} d^{n-1} x\,\sqrt{-\gamma}\,n^{\mu}\,\Big\{\xi_\sigma \nabla^\nu \nabla^\sigma\nabla_\nu \xi_\mu
-\xi_\mu \nabla^\nu\nabla^\sigma\nabla_\nu \xi_\sigma\nonumber\\
&&\qquad\quad+(\nabla_\mu \xi_\nu+\nabla_\nu \xi_\mu)(\Box\, \xi^\nu-\nabla^\nu\nabla^\sigma \xi_\sigma)
+R_{\mu\nu\rho\sigma}\xi^\rho\nabla^\sigma \xi^\nu\\
&&\qquad\quad+\frac{1}{2}\nabla_\mu\left[(\nabla^\nu \xi_\nu)^2-(\nabla^\nu \xi^\sigma)(\nabla_\nu \xi_\sigma)\right]
+2(\nabla^\nu \xi^\sigma)(\nabla_\nu\nabla_\sigma \xi_\mu-\nabla_\mu\nabla_\sigma \xi_\nu)\nonumber\\
&&\qquad\quad-\xi^\nu \nabla^\sigma\left[\xi_\mu (G_{\nu\sigma}+\Lambda g_{\nu\sigma})+\xi_\nu (G_{\mu\sigma}+\Lambda g_{\mu\sigma})-\xi_\sigma (G_{\mu\nu}+\Lambda g_{\mu\nu})\right] \Big\}\,.\nonumber
\end{eqnarray}

{\it If we assume\/} that  the background $g_{\mu\nu}$ satisfies the EFE, $G^{\mu\nu}+\Lambda g^{\mu\nu}=0$, then this second order action also reduces to a boundary term.  This is a consequence of gauge invariance of the second order action for $h$: assuming EFE for the background $g_{\mu\nu}$, $S^{(2)}[h]$ in \eqref{S0[h]+S1[h]+S2[h]} is invariant (changes only by a boundary term) under $h_{\mu\nu}\to h_{\mu\nu}+\nabla_{\mu}\xi_{\nu}+\nabla_{\nu}\xi_{\mu}$ and so the {\it ansatz\/} $h_{\mu\nu}=\nabla_{\mu}\xi_{\nu}+\nabla_{\nu}\xi_{\mu}$ into $S^{(2)}[h]$ is essentially pure gauge.  But if we  \emph{don't} assume that the background solves the EFE then \eqref{2nd order action for ksi --good writing} has a non-vanishing bulk effective action for $\xi$. Using the variational principle with $\delta \xi^\alpha$ and requiring that all boundary terms vanish as needed, the effective action up to second order yields a remarkably simple and beautiful equation of motion (EOM):
\be\label{EOM}
(G^{\mu\nu}+\Lambda g^{\mu\nu})(\xi_{\alpha;\mu\nu}-R_{\alpha\mu\nu\sigma}\xi^{\sigma})=0\,.
\ee
Here, as expected, we find that if the background $g_{\mu\nu}$ solves the EFE then the equation is vacuous; this is manifest from the first factor.  The second factor in \eqref{EOM} also has a simple meaning.  It is an equation satisfied by Killing vectors. For an exact Killing vector we have $\nabla_{\mu} \xi_\nu+\nabla_{\nu} \xi_\mu=0$ which, after taking a derivative and playing with the relevant equations a bit, implies $\xi_{\alpha;\mu\nu}-R_{\alpha\mu\nu\sigma}\xi^{\sigma}=0$ (see \cite{Wald:1984rg} for example). Thus for exact Killing vectors the EOM is again vacuous. This should have been expected since if $\xi$ is a Killing vector then our transformation $g_{\mu\nu}\to g_{\mu\nu}+\nabla_{\mu}\xi_{\nu}+\nabla_{\nu}\xi_{\mu}$ is not really a transformation and indeed $S^{(n)}[\xi]=0$ to all orders $n\geq 1$.    (We also recognize in the second factor Jacobi's equation for geodesic deviation.)

Thus we discover a memorable result: Our effective EOM \eqref{EOM} for \emph{approximate} solutions to Einstein's gravity and their \emph{approximate} symmetries is a contraction of two factors, each of which supports the existence of solutions required on general grounds.   (i) $G_{\mu\nu}+\Lambda g_{\mu\nu}=0\,$ is satisfied by exact solutions to Einstein's GR.   (ii) $\xi_{\alpha;\mu\nu}-R_{\alpha\mu\nu\sigma}\xi^{\sigma}=0\,$ is satisfied by exact Killing vectors.

Our EOM is satisfied in the \emph{asymptotic} limit by \emph{asymptotically} AdS$_3$ space-times \eqref{asymptotically AdS3 with single O}, \eqref{asymptotically AdS3 with relaxed Otr and Orphi} and their corresponding \emph{asymptotic} Killing vectors \eqref{asymptotic isometry vectors with single O}, \eqref{asymptotic isometry vectors with relaxed Otr and Orphi}, respectively. That is, (\ref{asymptotically AdS3 with single O}--\ref{asymptotic isometry vectors with single O}) and (\ref{asymptotically AdS3 with relaxed Otr and Orphi}--\ref{asymptotic isometry vectors with relaxed Otr and Orphi}) solve the equation:
\be\label{EOM in the limit r->infty}
\lim_{r\to\infty}(G^{\mu\nu}+\Lambda g^{\mu\nu})(\xi_{\alpha;\mu\nu}-R_{\alpha\mu\nu\sigma}\xi^{\sigma})=0\,.
\ee
The EOM vanishes fast enough that the integrated action / unit time can be arbitrarily small provided we begin with $r$ sufficiently large. Indeed, the $r$ integral in $S^{(2)}[\xi]_{bulk}$ in \eqref{2nd order action for ksi --good writing} is $dr\sqrt{-g}\,EOM_\alpha\,\xi^\alpha$ and in both cases $\sqrt{-g}={\cal O}(r)$ while $\xi^t={\cal O}(1), \xi^r={\cal O}(r), \xi^\phi={\cal O}(1)$, and so for the integrand: $S^{(2)}[\xi]_{bulk}\sim dr/r^3$ for (\ref{asymptotically AdS3 with single O}--\ref{asymptotic isometry vectors with single O}) and $S^{(2)}[\xi]_{bulk}\sim dr/r^2$ for (\ref{asymptotically AdS3 with relaxed Otr and Orphi}--\ref{asymptotic isometry vectors with relaxed Otr and Orphi}). Thus $S^{(2)}[\xi]_{bulk}$ can be arbitrarily small if one begins with $r$ sufficiently large. The boundary piece of $S^{(2)}[\xi]$ as well as the first order action $S^{(1)}[\xi]$ \eqref{boundary 1st order action for ksi} remain finite in both cases.

\section{The Virasoro Algebra for $AdS_3$}

We now assume power series expansion of the components of $\xi$ as follows:
\be\label{expansion}
\xi^\mu=\sum_n \xi^\mu_n (t,\phi)\, r^n\,.
\ee
We view the above as an expansion in $1/r\,$ (i.e. expansion around $r=\infty$) and we assume that each series truncates for some large $N$ onwards ($N$ may be different for each component).

We first obtain the Virasoro algebra generating vectors $\xi$ by requiring ``small'' asymptotic transformation of exact AdS$_3$.  ``Smallness'' is defined by requiring subleading Lie derivatives (along $\xi$) of AdS$_3$, finite first order effective action $S^{(1)}[\xi]$, and finite second order effective action $S^{(2)}[\xi]$.  Requiring that the ${\cal L}_\xi g_{\mu\nu}$ are subleading to the non-vanishing AdS$_3$ components $g_{\mu\nu}$ in \eqref{AdS3} leads to:
\be\label{subleading nonzero}
{\cal L}_\xi g_{tt}={\cal O}(r)\,, \quad {\cal L}_\xi g_{rr}={\cal O}(\frac{1}{r^3})\,, \quad {\cal L}_\xi g_{\phi\phi}={\cal O}(r)\,,
\ee
and we require that others remain finite:
\be\label{subleading finite}
{\cal L}_\xi g_{tr}={\cal O}(1)\,, \quad {\cal L}_\xi g_{t\phi}={\cal O}(1)\,, \quad {\cal L}_\xi g_{r\phi}={\cal O}(1)\,.
\ee
Using the expansion \eqref{expansion} the most general $\xi$'s satisfying these Lie derivative conditions are given by:
\begin{eqnarray}
\xi^t&=&l(T^++T^-)+\xi^t_{-1} (t,\phi)\frac{1}{r}+{\cal O}(\frac{1}{r^2})\,,\label{It}\\
\xi^r&=&-r\left({T^+}'+{T^-}'\right)+{\cal O}(1)\,,\label{Ir}\\
\xi^\phi&=&T^+-T^-+\xi^\phi_{-1} (t,\phi)\frac{1}{r}+{\cal O}(\frac{1}{r^2})\,,\label{Iphi}
\end{eqnarray}
with the additional constraint $l^2\xi^\phi_{-1,t}-\xi^t_{-1,\phi}=0$. This is already a very good start, as the leading terms have been fixed to the same ones that appeared in \eqref{asymptotic isometry vectors with single O} and \eqref{asymptotic isometry vectors with relaxed Otr and Orphi}. Note, however, that in order to close the Virasoro algebra at leading order in $1/r$ we also want the next to leading terms to vanish.

The first order effective action $S^{(1)}[\xi]$ \eqref{boundary 1st order action for ksi} is an integral over the boundary at $r=\infty$. We require this to be finite, in the sense that the integrand is finite everywhere on the boundary.  (We are not going to be concerned with divergences due to infinite time range.) We find that for $S^{(1)}[\xi]$ to remain finite the leading order terms in (\ref{It}--\ref{Iphi}) are already related as needed, but at second to leading order the arbitrary ${\cal O}(1)$ piece in $\xi^r$ in \eqref{Ir} must satisfy $\xi^r_0=-(\xi^t_{-1,t}+\xi^\phi_{-1,\phi})/4$.

Since AdS$_3$ is an exact solution of EFE, the second order effective action $S^{(2)}[\xi]$ \eqref{2nd order action for ksi --good writing} is also a boundary term (i.e. an integral over the boundary at $r=\infty$). We require $S^{(2)}[\xi]$ to be finite in the same sense as before.  We find that this imposes $\xi^t_{-1}=\pm l\,\xi^\phi_{-1}$ at leading order and eventually, if we allow for arbitrary subdominant terms $\xi^t_{-2}$ and  $\xi^\phi_{-2}$, leads to $\xi^t_{-1}=\xi^\phi_{-1}=0$.  We therefore arrive at:
\begin{eqnarray}
\xi^t&=&l(T^++T^-)+{\cal O}(\frac{1}{r^2})\,,\nonumber\\
\xi^r&=&-r\left({T^+}'+{T^-}'\right)+{\cal O}(\frac{1}{r})\,,\label{new ksis}\\
\xi^\phi&=&T^+-T^-+{\cal O}(\frac{1}{r^2})\,.\nonumber
\end{eqnarray}
The $\xi$'s in \eqref{new ksis} are our final vector fields for small asymptotic transformations of AdS$_3$. They define proper perturbations for  AdS$_3$, for which both $S^{(1)}[\xi]$ and $S^{(2)}[\xi]$ are finite.
They obey the classical, centerless Virasoro algebra \eqref{centerless Virasoro} at leading order in $1/r$.  Thus the  Virasoro algebra emerges in the asymptotic transformations of AdS$_3$ independent of any {\it a priori\/} choice of boundary conditions.

Since the asymptotics of AdS$_3$ are governed by the conformal group in two dimensions, we are encouraged to consider  asymptotic \emph{conformal} Killing vectors of AdS$_3$ directly.  We now show that asymptotic conformal Killing vectors of AdS$_3$ which leave the first order effective action finite, also end up taking the form \eqref{new ksis}. Assume the vector field $\xi$ satisfies the conformal Killing equation for AdS$_3$,
\be\label{ckv}
\nabla_\mu\xi_\nu+\nabla_\nu\xi_\mu=\frac{2}{3}g_{\mu\nu}\nabla_\sigma\xi^\sigma\,,
\ee
in the asymptotic limit $r\to\infty$. Starting from equation \eqref{1st order action for ksi} and using the conformal Killing equation \eqref{ckv} on the boundary integral (where $r=\infty$), we may simplify $S^{(1)}[\xi]$ for AdS$_3$ to the following:
\be\label{S1[xi]-ckv}
S^{(1)}[\xi]=-\frac{4}{3}\int_{\partial M} d^{2} x\,\sqrt{-\gamma}\,n^{\mu} \nabla_{\mu}\nabla_{\sigma}\xi^{\sigma}\,.
\ee
We find that the most general $\xi$'s satisfying the conformal Killing equation for AdS$_3$ \eqref{ckv} in the limit $r\to\infty$ while also maintaining a finite first order effective action $S^{(1)}[\xi]$ \eqref{S1[xi]-ckv}, in the sense that the integrand is finite everywhere on the boundary $r=\infty$, are given by:
\begin{eqnarray}
\xi^t&=&l(T^++T^-)+\xi^t_{-2}(t,\phi)\frac{1}{r^2}+{\cal O}(\frac{1}{r^3})\,,\label{final-ckv-t}\\
\xi^r&=&-r\left({T^+}'+{T^-}'\right)+\xi^r_{-1}(t,\phi)\frac{1}{r}+{\cal O}(\frac{1}{r^2})\,,\label{final-ckv-r}\\
\xi^\phi&=&T^+-T^-+\xi^\phi_{-2}(t,\phi)\frac{1}{r^2}+{\cal O}(\frac{1}{r^3})\,,\label{final-ckv-phi}
\end{eqnarray}
with $\xi^t_{-2}, \xi^\phi_{-2}, \xi^r_{-1}$ satisfying the following equations:
\begin{eqnarray}
&&2\xi^\phi_{-2,\phi}-\xi^t_{-2,t}+3\xi^r_{-1}=0\,,\label{addl subleading rln 1}\\
&&l^2\left({T^+}'+{T^-}'\right)+\xi^t_{-2,t}-\xi^\phi_{-2,\phi}=0\,,\label{addl subleading rln 2}\\
&&l^3\left({T^+}'-{T^-}'\right)+\xi^t_{-2,\phi}-l^2\xi^\phi_{-2,t}=0\,.\label{addl subleading rln 3}
\end{eqnarray}
Since the equations (\ref{addl subleading rln 1}--\ref{addl subleading rln 3}) may be solved for $\xi^t_{-2}, \xi^\phi_{-2}, \xi^r_{-1}$, without imposing any constraints on $T^+$ and $T^-$ (\emph{e.g.} $\xi^t_{-2}=0, \xi^\phi_{-2}=l^2(T^+-T^-), \xi^r_{-1}=-2l^2({T^+}'+{T^-}')/3$), we find that our final $\xi$'s (\ref{final-ckv-t}--\ref{addl subleading rln 3}) are again of the form \eqref{new ksis}. Hence they form the Virasoro algebra \eqref{centerless Virasoro} at leading order in $1/r$.

We will see in the next section that the vectors \eqref{new ksis} define asymptotic symmetries of some new asymptotically AdS$_3$ space-times.   Since we have derived the Virasoro algebra independently  we may proceed to the canonical generators, arriving at the same central charge $c=3l/(2G)$ at the Dirac bracket level and in the end the same quantum Virasoro algebra of a two-dimensional CFT.

\section{New asymptotically $AdS_3$ spaces}

Having established status for the vectors \eqref{new ksis} using the effective action approach in the previous section let us here attempt to embed them back into the context of an asymptotic symmetry group. Define \emph{new asymptotically AdS$_3$ space-times} by perturbing the exact AdS$_3$ metric \eqref{AdS3} using our $\xi$'s \eqref{new ksis}:
\begin{equation}\label{new asymptotically AdS3 with single O}
\begin{gathered}
g_{tt}=-\frac{r^2}{l^2}+{\cal O}(1)\,, \quad g_{tr}={\cal O}(\frac{1}{r})\,,\quad g_{t\phi}={\cal O}(1)\,,\\
g_{rr}=\frac{l^2}{r^2}+{\cal O}(\frac{1}{r^4})\,,\quad g_{r\phi}={\cal O}(\frac{1}{r})\,,\quad g_{\phi\phi}=r^2+{\cal O}(1)\,.
\end{gathered}
\end{equation}
Note that these are relaxed compared to both Brown-Henneaux \eqref{asymptotically AdS3 with single O} and our \eqref{asymptotically AdS3 with relaxed Otr and Orphi}. It is easy to show that the asymptotic symmetries of \eqref{new asymptotically AdS3 with single O} are actually all given by our $\xi$'s \eqref{new ksis}. Thus, provided the corresponding charges are still well defined, the ASG of AdS$_3$ corresponding to the boundary conditions \eqref{new asymptotically AdS3 with single O} is again the conformal group in two dimensions. Indeed, one finds that the charges $Q[\xi]$ corresponding to (\ref{new ksis}, \ref{new asymptotically AdS3 with single O}) are again finite. The central charge of the Virasoro algebra at the Dirac bracket level is again $c=3l/(2G)$ and it may be used to calculate the BTZ entropy via the Cardy formula as usual. We also note that, as with all cases of asymptotically AdS$_3$ space-times and corresponding asymptotic symmetries considered in this Letter, the $r\to\infty$ limiting EOM \eqref{EOM in the limit r->infty} is satisfied by the set (\ref{new ksis}, \ref{new asymptotically AdS3 with single O}) fast enough that the integrated action / unit time can be arbitrarily small if one begins with $r$ sufficiently large. Indeed, we have for (\ref{new ksis}, \ref{new asymptotically AdS3 with single O}): $S^{(2)}[\xi]_{bulk}\sim dr/r^3$ and the boundary piece of $S^{(2)}[\xi]$ as well as the first order action $S^{(1)}[\xi]$ are finite.

\section{Conclusion and Discussion}

To date the Virasoro algebra associated with the boundary of AdS$_3$ was known to arise in the context of an asymptotic symmetry group (ASG) corresponding to certain boundary conditions imposed by Brown and Henneaux. Although these boundary conditions play a key role in obtaining the Virasoro algebra \emph{a la} Brown-Henneaux, they are not entirely dictated by the theory. We have bridged this gap by \emph{deriving} boundary conditions from finiteness constraints on the effective action(s) of GR calculated in this Letter. The boundary conditions so derived are relaxed compared to Brown-Henneaux but the corresponding ASG and its central charge are unaltered: It is again the conformal group in two dimensions with $c=3l/(2G)$.

However, we need not postulate boundary conditions and an ASG {\it a priori}. As emphasized throughout this Letter, the emergence of the Virasoro algebra in the asymptotics of AdS$_3$ can be grounded in the physical requirement of finite action.

We have mentioned in the introduction that the two-dimensional conformal symmetry in the boundary of AdS$_3$ is relevant for calculating entropy of the BTZ black hole. We have also mentioned that for appropriate boundary conditions the ASG of the near horizon metric of extremal Kerr (NHEK) also contains the conformal group with a central charge that gives entropy of extremal Kerr. One may apply our effective action approach for this case too and we shall report the results elsewhere soon.

Two-dimensional conformal symmetry appears in the physics of realistic Kerr black holes as well.  Recently scattering amplitudes by near-extremal Kerr were obtained from certain correlators of two-dimensional CFT \cite{Bredberg:2009pv}. Moreover, a new hidden two-dimensional conformal symmetry in the low frequency scalar wave equation of generic Kerr has also been found very recently in \cite{Castro:2010fd}. These papers signal a duality between Kerr black holes and two-dimensional conformal field theories that apparently cannot be captured by an ASG.  It may be possible to obtain a unifying realization of all conformal symmetry associated with the Kerr metric by considering small action {\it and energy\/} constraints on relevant transformations in the Kerr geometry.

\acknowledgements
A.~P.~P. would like to thank Sean Robinson for helpful discussions.

\end{document}